\documentclass[]{spie}  %>>> use for US letter paper
\usepackage[]{graphicx}

\title{Using ACIS on the Chandra X-ray Observatory as a \\ particle radiation monitor} 

\author{C. E. Grant\supit{a}, B. LaMarr\supit{a}, M. W. Bautz\supit{a} and S. L. O'Dell\supit{b}
\skiplinehalf
\supit{a}Kavli Institute for Astrophysics and Space Research, Massachusetts Institute of Technology, Cambridge, Massachusetts, USA; \\
\supit{b}NASA Marshall Space Flight Center, Huntsville, Alabama, USA
}

\authorinfo{Further author information: (Send correspondence to C.E.G.)\\C.E.G.: E-mail: cgrant@space.mit.edu}

\begin{document} 
  \maketitle 

\begin{abstract}
The Advanced CCD Imaging Spectrometer (ACIS) is one of two focal-plane instruments on the Chandra X-ray Observatory.  During initial radiation-belt passes, the exposed ACIS suffered significant radiation damage from trapped soft protons scattering off the x-ray telescope's mirrors.  The primary effect of this damage was to increase the charge-transfer inefficiency (CTI) of the ACIS 8 front-illuminated CCDs.  Subsequently, the Chandra team implemented procedures to remove the ACIS from the telescope's focus during high-radiation events: planned protection during radiation-belt transits; autonomous protection triggered by an on-board radiation monitor; and manual intervention based upon assessment of space-weather conditions. However, as Chandra's multilayer insulation ages, elevated temperatures have reduced the effectiveness of the on-board radiation monitor for autonomous protection.  Here we investigate using the ACIS CCDs themselves as a radiation monitor.  We explore the 10-year database to evaluate the CCDs' response to particle radiation and to compare this response with other radiation data and environment models.

\end{abstract}


\section{INTRODUCTION}
\label{sec:intro}

The Chandra X-ray Observatory, the third of NASA's great observatories in space, was launched just past midnight on July 23, 1999, aboard the space shuttle {\it Columbia}\cite{cha2}.  After a series of orbital maneuvers Chandra reached its final, highly elliptical, orbit.  Chandra's orbit, with a perigee of 10,000~km, an apogee of 140,000~km and an initial inclination of 28.5$^\circ$, transits a wide range of particle environments, from the radiation belts at closest approach through the magnetosphere and magnetopause and past the bow shock into the solar wind.

The Advanced CCD Imaging Spectrometer (ACIS), one of two focal plane science instruments on Chandra, utilizes frame-transfer charge-coupled devices (CCDs) of two types, front- and back-illuminated (FI and BI)\cite{acis}.  Soon after launch it was discovered that the FI CCDs had suffered radiation damage from exposure to soft protons scattered off the Observatory's grazing-incidence optics during passages through the Earth's radiation belts\cite{gyp00}.  Since mid-September 1999, ACIS has been protected during radiation belt passages and there is an ongoing effort to prevent further damage and to develop hardware and software strategies to mitigate the effects of charge transfer inefficiency on data analysis\cite{odell}.

Our primary measure of radiation damage on the CCDs is charge transfer inefficiency (CTI).  The eight front-illuminated CCDs had essentially no CTI before launch, but are strongly sensitive to radiation damage from low energy protons ($\sim$100~keV) which preferentially create traps in the buried transfer channel.  The framestore covers are thick enough to stop this radiation, so the initial damage was limited to the imaging area of the FI CCDs.  Radiation damage from low-energy protons is now minimized by moving the ACIS detector away from the aimpoint of the observatory during passages through the Earth's particle belts.  Continuing exposure to both low and high energy particles over the lifetime of the mission slowly degrades the CTI further.\cite{odell,ctitrend}  The two back-illuminated CCDs (ACIS-S1,S3) suffered damage during the manufacturing process and exhibit CTI in both the imaging and framestore areas and the serial transfer array, but are less sensitive to the low energy particles which damage the FI CCDs because they cannot reach the much deeper transfer channel. 

Since early in the Chandra mission, procedures have been implemented which protect the focal plane instruments during times of high radiation.  ACIS is translated out of the focal plane, providing protection against soft protons, and is powered off.  Three types of procedures are in place; planned protection during radiation-belt transits, autonomous protection triggered by the on-board radiation monitor, and manual intervention based upon assessment of space-weather conditions.  The Chandra weekly command load includes automatic scheduled safing of the focal-plane instruments during radiation belt passages.  The timing of radiation belt ingress and egress are determined using the standard AP8/AE8 environment with a small additional pad time to protect against temporal variations.  Solar storms are detected by either the on-board radiation monitor or by ground operations monitoring of various space weather measures, such as from the ACE and GOES spacecraft and the Kp index.  The on-board radiation monitor cannot detect protons at hundreds of keV, which are the most damaging to ACIS, so on-board protection is supplemented by other measures of the radiation environment. 

The Electron, Proton, Helium INstrument (EPHIN) is a particle detector on-board the Chandra spacecraft used to monitor the local particle radiation environment.  It is sensitive to electrons in the energy range 150 keV - 50 MeV and protons from 5 - 49 MeV.  Chandra-EPHIN is very similar to the EPHIN detector onboard SOHO.  Until December 2008, EPHIN rates in three channels were monitored by the spacecraft computer which can command radiation shutdowns during solar storms.  The monitored channels were P4, sensitive to protons with 5.0-8.3 MeV, P41, sensitive to protons with 41-53 MeV and E1300, sensitive to electrons with 2.64-6.18 MeV.  As the spacecraft insulation has aged and degraded, thermal control of some subsystems, including EPHIN, has become more difficult.  Elevated EPHIN temperatures cause anomalous noise, which in some EPHIN channels can be significant and occasionally dominate the signal.  For more on EPHIN degradation see Ref.~\citenum{odell}.  To prevent against false triggers due to EPHIN noise, EPHIN was reconfigured in December 2008.  The spacecraft computer now monitors two EPHIN channels which do not distinguish between particle species.  Concern about the effectiveness of EPHIN going into the future motivates looking for other on-board measures of the radiation environment.  One such measure, explored elsewhere, uses the other Chandra focal plane instrument, the High Resolution Camera (HRC)\cite{odell}.  A second measure, examined here, is using ACIS itself.

In this paper, we present an initial exploration into the effectiveness of using ACIS as its own particle monitor.  Section~\ref{sec:acis} describes the type of ACIS data we choose to examine and the radiation event data is outlined in section~\ref{sec:rad}.  The ACIS and radiation data are compared in Section~\ref{sec:anal}, while Section~\ref{sec:concl} summarizes our results and suggests directions for future work.

\section{ACIS DATA}
\label{sec:acis}

We are somewhat limited in what type of information ACIS can collect for radiation monitoring purposes.  While the ACIS flight software is quite flexible, the primary purpose of ACIS is always to collect astrophysical data so any particle monitoring processes need to be secondary in using the available on-board resources.  ACIS observations are done in a variety of configurations with different frametimes, subarrays, windowing, energy filters and CCDs to meet various science goals.  Small changes in science operations may be possible with little to no impact on the science.  For example, using a particular CCD as a particle monitor would require it to be turned on during all observations, regardless of the observer's needs, but for most scientific purposes the impact would be minimal.

To reduce the telemetry required for standard operations, ACIS processes raw data frames on-board to find potential X-ray events and filter out charged particle events.  The event lists are then telemetered to the ground, along with exposure records for each raw frame.  The exposure records include the number of pixels in the frame above the event threshold, the number of events found, and the number of events discarded by various filters.  As a first test of a potential ACIS radiation monitor, we choose to use this exposure record information and more specifically, the number of pixels above the event threshold, the threshold crossing rate.  Unlike the accepted event rate, which is strongly affected by X-ray sources, or the discarded event rates, which are dependent on the observing mode, the threshold crossing rate is less sensitive to sources or observing mode and more sensitive to the particle rate.  

One difficulty of using actual ACIS data is that ACIS, by the design of the radiation protection plan, is turned off when the radiation environment is believed to be high.  The ACIS data is artificially truncated at a particular particle rate, limiting our ability to correlate with other radiation measures.  A possible strength of using the real telemetry is that it reflects the true heterogeneity of ACIS observing modes and X-ray source types.  Any automated radiation monitor must be robust against false triggers due to high X-ray source rates or unusual ACIS science modes.

As discussed above, we have chosen to use the threshold crossing rate in an initial test of the possibility of using ACIS as its own radiation monitor.  We have collected all the threshold crossing data from the entire lifetime of Chandra from August 1999 through May 2010.  The raw data format is counts per frame, so to correct for the variety of observing modes with different frame times, we have taken five minute average rates in counts per second.  Similarly, we have corrected the data to account for modes where only a portion of the CCD is accumulating counts, so the final data product is in units of counts/sec/CCD.  We make no corrections for counts from any X-ray sources in the field of view.  To ensure maximal time coverage, we concentrate on data from the BI CCD ACIS-S3 and FI CCD ACIS-I3, which contain the telescope focal point in the spectroscopic and imaging configurations.  All ACIS observations utilize at least one of these CCDs, and often both.  At the present time we are not considering data taken when ACIS is out of the focal plane observing its radioactive calibration source.  Radiation protection monitoring is not active during these times, so an ACIS-based monitor would also not be required to accurately sense the radiation environment in that configuration.

\begin{figure}
\vspace{4.8in}
\includegraphics{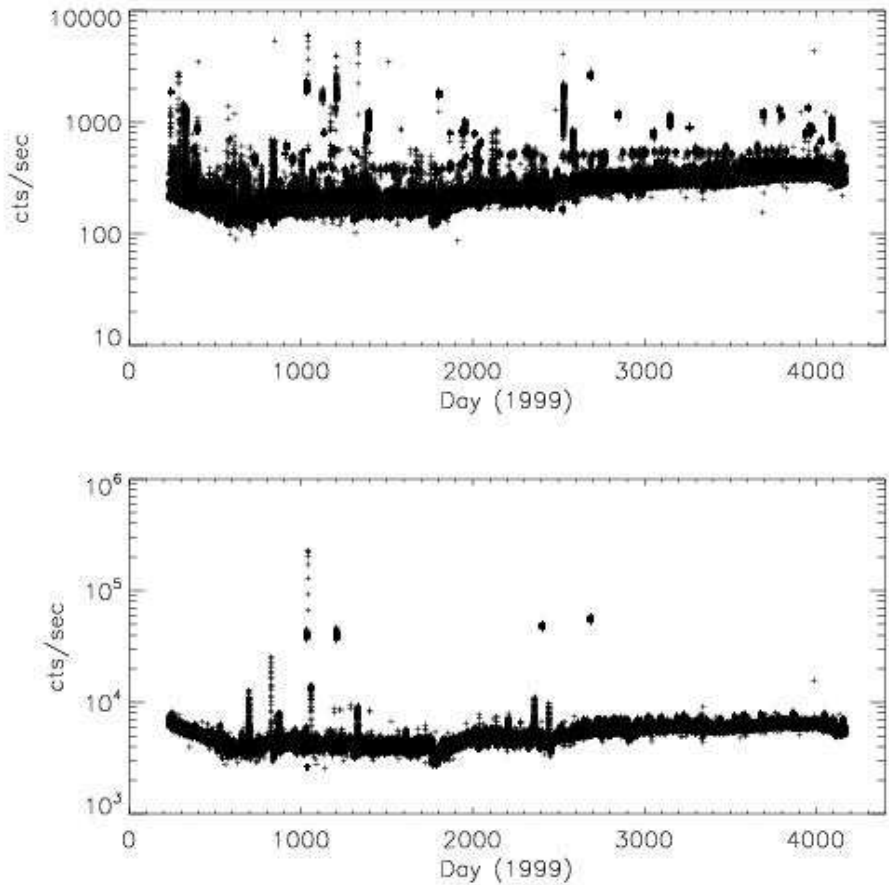}
\caption{Threshold crossing rate as a function of time over the ten year history of Chandra for the back-illuminated ACIS-S3 (top) and the front-illuminated ACIS-I3 (bottom).  Each data point is a five minute average.}
\label{fig:threshall}
\end{figure}

This ten year history of threshold crossing data while ACIS is the focal plane instrument is shown in Figure~\ref{fig:threshall}.  The two panels demonstrate the different response of the back- and front-illuminated CCDs to charged particles.  Although the active depleted region of both CCDs has a similar depth, the BI CCD has been thinned, while the FI CCD has a substantial field-free region.  Because of this, particle events produce much larger charge blooms on FI CCDs than on BI CCDs, and so a single particle event will produce many more threshold crossings on an FI CCD.  With the much lower particle induced threshold crossing rate on the BI CCD, brighter X-ray sources, particularly bright diffuse objects are apparent above the quiescent background.  The vertical streaks are potentially increases in particle rates associated with solar storms.  Figure~\ref{fig:threshall} also shows the variability of the quiescent particle background which is anti-correlated with the solar cycle.

\section{RADIATION EVENT DATA}
\label{sec:rad}

In its ten year lifetime, Chandra has shutdown sixty-five times due to a high radiation environment.  Because of reduced activity during solar minimum, there have been no radiation shutdowns since 2006.  In twenty-one of those cases, there is no ACIS event data of the sky in the hour previous to the shutdown.  Either ACIS was taking data of its calibration source, out of the focal plane, or ACIS was not taking data at all.  In addition, we are primarily interested in the radiation shutdowns that were triggered autonomously by the on-board radiation monitor rather than from ground commanding.  This leaves just twenty-seven shutdowns to examine.

An additional factor is that some of the autonomous radiation shutdowns immediately precede a scheduled radiation-zone shutdown.  These are due to electron-flux spikes that present no danger to the instruments\cite{odell}.  The frequency of these events was reduced by increasing the required number of above-threshold samples used by the on-board monitoring algorithm.  As these events are related to the Earth's radiation belts and not to solar storms, we expect them to exhibit a different particle spectrum and thus may also induce a different response in ACIS.  There are ten of these radiation belt events.

\begin{figure}
\vspace{7.3in}
\includegraphics{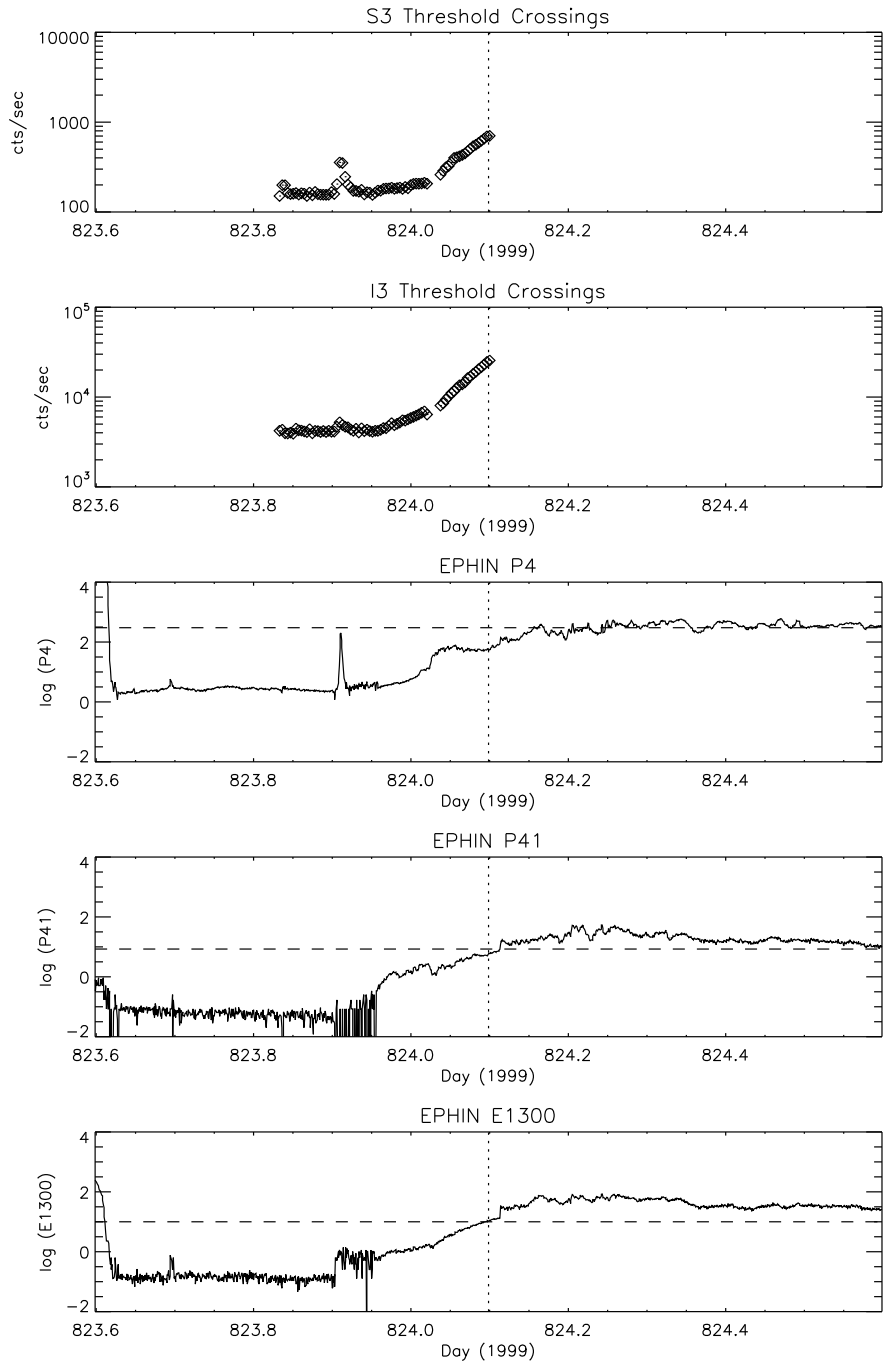}
\caption{Lightcurves during a solar storm on day 93 of 2001.  The top two panels are the threshold crossing rate on the BI CCD S3 and the FI CCD I3.  The bottom three panels are the rates measured by the on-board EPHIN detector in three channels: P4 (protons with 5.0-8.3 MeV), P41 (protons with 41-53 MeV) and E1300 (electrons with 2.64-6.18 MeV) in counts/cm$^2$/sec/sr/MeV.  The horizontal dashed lines indicate the trigger threshold for an autonomous shutdown of the spacecraft.  The vertical dotted lines are the time of the autonomous shutdown.  The initially high EPHIN rates are due to a radiation belt passage.}
\label{fig:ephinplot}
\end{figure}

\begin{figure}
\vspace{7.3in}
\includegraphics{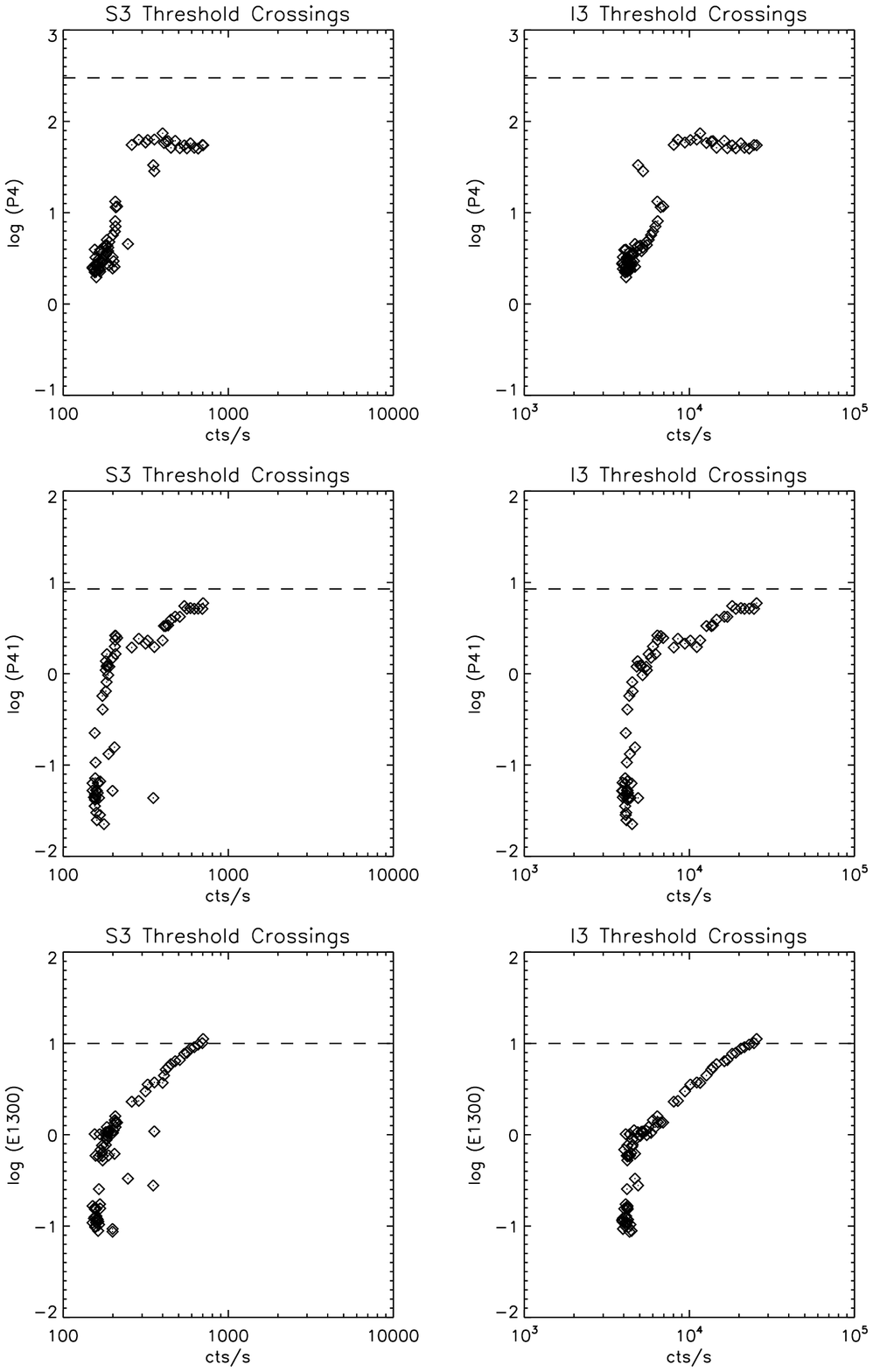}
\caption{Comparison of the threshold crossing rates on ACIS with the rates in the three EPHIN channels for the solar storm on day 93 of 2001.  The data is identical to that in Figure~\ref{fig:ephinplot}.  The horizontal dashed lines indicate the trigger threshold for an autonomous shutdown of the spacecraft.}
\label{fig:ephincorr}
\end{figure}

For each of the twenty-seven shutdowns that were autonomous, we have made lightcurves of the threshold crossing rate on the BI CCD ACIS-S3 and the FI CCD ACIS-I3.  We then compare these lightcurves to the particle rates measured by the on-board EPHIN detector in the P4 (protons with 5.0-8.3 MeV), P41 (protons with 41-53 MeV) and E1300 (electrons with 2.64-6.18 MeV) channels, which are the channels used for on-board radiation monitoring.  Figure~\ref{fig:ephinplot} is an example of these lightcurves for an autonomous radiation shutdown that took place on day 93 of 2001.  This is a particularly good case where both types of CCDs were taking data during a rapid rise in the particle rates as measured by EPHIN and which is also seen by ACIS.  Figure~\ref{fig:ephincorr} shows the correlation of ACIS and EPHIN rates for the same radiation shutdown.  While all three EPHIN channels show an increase in rates, the shape and timing of the rise is slightly different.  The ACIS rates seem to best match the increase seen in the E1300 channel and do not match the P4 channel very well at all.

\section{ANALYSIS}
\label{sec:anal}

Some initial observations can be made after inspection of the ACIS and EPHIN lightcurves.  Most, but not all, of the radiation shutdowns exhibit some increase in the threshold crossing rates on ACIS, although the size of the increase is sometimes small.  In a few cases, like the one shown in Figure~\ref{fig:ephinplot} and \ref{fig:ephincorr}, the increase is substantial and well correlated with at least one EPHIN channel.  Whether the increase in rates is clear to the eye in the ACIS data is often related to the shape of the particle increase - a single rapid increase is much easier to spot than a more gradual one.  If the increase is too rapid, radiation protection procedures may shut off the instruments before much if any data is obtained.

In the five cases in which the High-Energy Transmission Grating (HETG) is inserted, none show any change in ACIS rates at all.  This is not entirely surprising as the HETG grating facets are supported by a thin polyimide structure (5500 or 9800 \AA) which absorbs and Rutherford scatters some of the particle radiation.  We do not have any radiation shutdown data with the Low-Energy Transmission Grating (LETG) but as the grating facets are free-standing, we do not expect the same level of radiation protection.  This does not necessarily imply that ACIS cannot sense particle radiation at all while the HETG is inserted.  On the contrary, the strongest radiation events can be seen in the threshold crossing rates while ACIS is out of the focal plane altogether, with much more shielding than just the HETG can provide.  It is plausible however that the sensitivity to lower fluxes and lower energies may be reduced while the HETG is inserted.

\begin{figure}
\vspace{7.3in}
\includegraphics{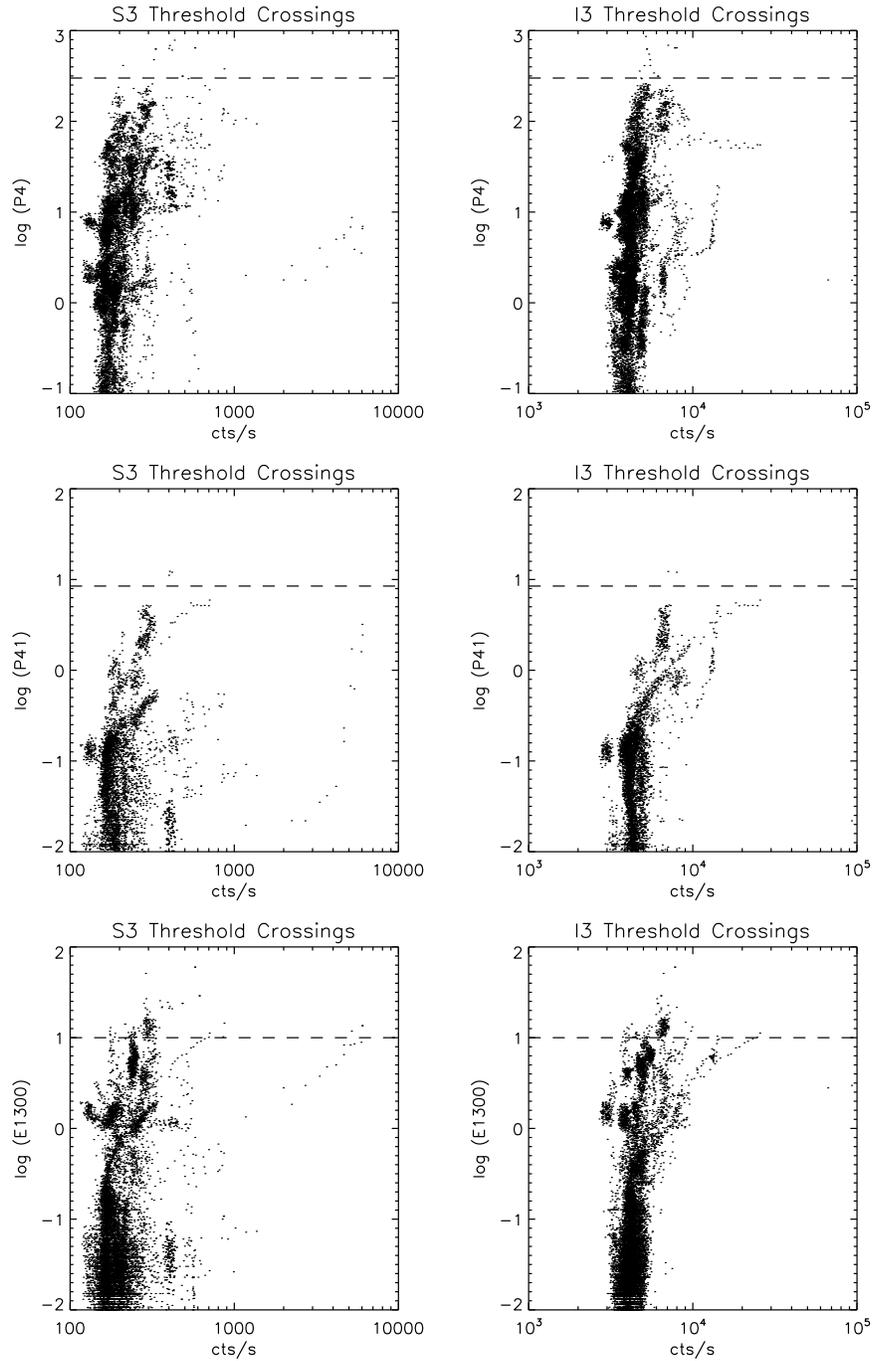}
\caption{Comparison of the threshold crossing rates on ACIS with the rates in the three EPHIN channels for all data within two days of a radiation shutdown.  The horizontal dashed lines indicate the trigger threshold for an autonomous shutdown of the spacecraft.  This plot includes data within two days of all radiation shutdowns in which ACIS is in the focal plane and the gratings are not inserted, including those commanded from the ground.}
\label{fig:bigcorr}
\end{figure}

Figure~\ref{fig:bigcorr} shows the correspondence between ACIS threshold crossing rates and EPHIN for all sixty radiation shutdowns with ACIS in the focal plane without the gratings inserted, including those shutdowns commanded from the ground.  While there certainly is some correlation between the ACIS and EPHIN rates, it does not appear to be simple or consistent between radiation events.  Some individual events can be seen snaking up and to the right, all with different slopes.  The S3 rates show the most low count rate scatter, most likely due to X-ray sources.  These radiation events are all different in particle species and energies, which is partly the cause of the differing response by ACIS.  In addition, EPHIN is not sensitive to the lowest energy particles, such as protons below 5~MeV, while ACIS is, and these may also help explain the cases where ACIS rates increase but are not well correlated with the EPHIN increase.

\begin{figure}
\vspace{4.9in}
\includegraphics{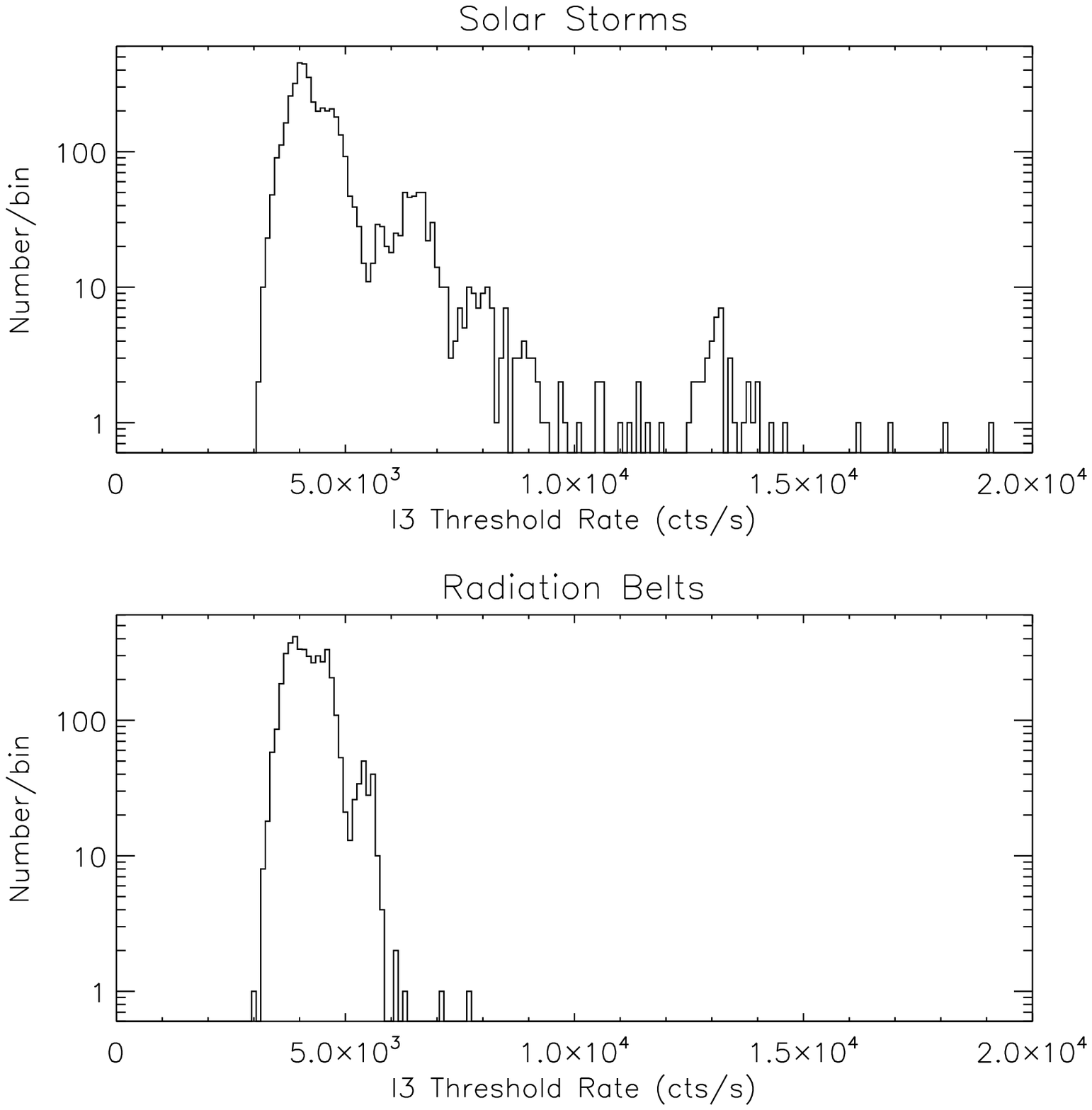}
\caption{Histogram of threshold crossing rates for ACIS-I3 during autonomous radiation shutdowns due to solar storms (top) and radiation belts (bottom).  The solar storm histogram has many more high count rate outliers than the one for the radiation belt events.}
\label{fig:threshhist}
\end{figure}

A demonstration of the response to different particle spectra is shown in Figure~\ref{fig:threshhist} which shows histograms of threshold crossing rates during autonomous radiation shutdowns divided into two categories.  The top histogram is shutdowns due to solar events, while the bottom histogram is shutdowns immediately prior to radiation belt entry.  The threshold crossing distribution for the solar storm events has many more high count rate outliers than the one for the radiation belts.

It is clear based on the previous analysis that the ACIS response to particle radiation is not entirely consistent with the behavior of the EPHIN channels used for radiation monitoring.  An ACIS-based radiation trigger, at least one utilizing threshold crossing rates, might not activate for all the radiation events that an EPHIN trigger would.  Given that there are other methods of detecting high radiation environments, including monitoring other measures from the ground and the HRC anti-coincidence rates on-board, partial protection from ACIS monitoring could still have some value.  An ACIS-based radiation trigger, though, has the potential for false triggering due to bright X-ray sources.  

To test both the effectiveness of an ACIS radiation trigger and against the possibility of false triggers, we applied a trigger level of an I3 threshold crossing rate of $9 \times 10^3$ cts/s for at least 10 minutes to the entire ten year dataset.  The I3 CCD was chosen because of the cleaner lightcurve due to the much stronger particle radiation signal.  The trigger threshold level was chosen somewhat arbitrarily by examining the time evolution and scatter of the quiescent level in Figure~\ref{fig:threshall}.

This test trigger would have shutdown the spacecraft seven times.  Four of these are within a few hours of a real EPHIN shutdown due to a solar storm.  Two precede an EPHIN shutdown by about two days and are due to an initial flare in a larger solar event.  One was due to a real solar storm that came close to but did not reach the EPHIN trigger levels.  There were no false triggers, in the sense that these all represent real times of high radiation levels.  It is unknown whether ACIS would have reached the trigger level within a few hours after the EPHIN shutdown, but in many cases it seems plausible that it could, so this could be considered a lower limit on the number of ACIS-based radiation triggers.

To gage the effectiveness of the ACIS-based trigger, we need to know how many of the real EPHIN solar storm shutdowns occurred while I3 was taking data and thus had the potential to be detected by an ACIS-I3 trigger.  There are twelve total autonomous radiation shutdowns from solar storms in which I3 is taking data.  The ACIS trigger would have caught four of them, six if the two initial flares are included, so at least a third of the EPHIN triggers or as many as half were detected.

\section{RESULTS}
\label{sec:concl}

We have examined one possible metric, the threshold crossing rate, that could be used as part of an ACIS-based particle radiation monitor.  The correlation with the monitored EPHIN particle channels is not consistent between radiation events or particularly strong overall, most likely due to different distributions of particle species and spectra in each event and the different sensitivity of each instrument.  As there are radiation events in which ACIS exhibits a strong response, we tested the effectiveness of a potential ACIS-based radiation trigger and found that at least half the radiation events detected by EPHIN would also have been found by ACIS with no false triggers.  Given that an ACIS monitor would likely be run in conjunction with other measures, such as the HRC anti-coincidence rate, ground monitoring, and even a less-effective EPHIN, this partial radiation protection may be sufficient.

Before implementing such a radiation monitor, more analysis would be required.  The current study compares ACIS threshold crossing rates to EPHIN during time periods adjacent to radiation shutdowns.  A more complete study would require comparison of the EPHIN rates over the entire lifetime of Chandra.  We have confined ourselves to a particular ACIS measure, the threshold crossing rate, but any number of measures can be imagined.  Exploration of a few different options, to confirm that our assumptions about the utility of threshold crossings is important.  Finally, as each radiation event is unique in flux and spectrum, it may also be interesting to examine in more detail what characteristics of the events (energy spectra, particle species, rise/decay-time) can be linked to stronger or weaker ACIS response.

\acknowledgments
We would like to thank Paul Plucinsky, Scott Wolk, the Chandra Science Operations team, and the Chandra Project Science team for many years of fruitful collaboration and constant vigilance in managing ACIS radiation damage.  This work was supported by NASA contracts NAS 8-37716 and NAS 8-38252. 

\bibliography{article}

\begin{thebibliography}{1}

\bibitem{cha2}
{Weisskopf}, M.~C., {Brinkman}, B., {Canizares}, C., {Garmire}, G., {Murray},
  S., and {Van Speybroeck}, L.~P., ``{An Overview of the Performance and
  Scientific Results from the Chandra X-Ray Observatory},'' {\em Pub. of the
  Astron. Society of the Pacific}~{\bf 114},  1--24 (2002).

\bibitem{acis}
{Garmire}, G.~P., {Bautz}, M.~W., {Ford}, P.~G., {Nousek}, J.~A., and {Ricker
  Jr.}, G.~R., ``Advanced {CCD} {I}maging {S}pectrometer ({ACIS}) instrument on
  the {C}handra {X}-ray {O}bservatory,'' in [{\em X-Ray and Gamma-Ray
  Telescopes and Instruments for Astronomy}{\nolinebreak\hspace{0.1em}]},
  Truemper, J.~E. and Tanabaum, H.~D., eds., {\em Proc. SPIE} {\bf 4851},
  28--44 (2003).

\bibitem{gyp00}
{Prigozhin}, G.~Y., {Kissel}, S.~E., {Bautz}, M.~W., {Grant}, C., {LaMarr}, B.,
  {Foster}, R.~F., and {Ricker}, G.~R., ``Characterization of the radiation
  damage in the {C}handra x-ray {CCD}s,'' in [{\em X-ray and Gamma-Ray
  Instrumentation for Astronomy XI}{\nolinebreak\hspace{0.1em}]},  Flanagan,
  K.~A. and Siegmund, O.~H., eds., {\em Proc. SPIE} {\bf 4140},  123--134
  (2000).

\bibitem{odell}
{O'Dell}, S.~L. et~al., ``Managing radiation degradation of {CCD}s on the
  {C}handra {X}-ray {O}bservatory {III},'' in [{\em UV, X-Ray, and Gamma-Ray
  Space Instrumentation for Astronomy XV}{\nolinebreak\hspace{0.1em}]},
  Siegmund, O.~H.~W., ed., {\em Proc. SPIE} {\bf 6686},  668603--1--668603--12
  (2007).

\bibitem{ctitrend}
{Grant}, C.~E., {Bautz}, M.~W., {Kissel}, S.~M., {LaMarr}, B., and {Prigozhin},
  G.~Y., ``Long-term trends in radiation damage of {C}handra x-ray {CCD}s,'' in
  [{\em UV, X-Ray, and Gamma-Ray Space Instrumentation for Astronomy
  XIV}{\nolinebreak\hspace{0.1em}]},  Siegmund, O.~H.~W., ed., {\em Proc. SPIE}
  {\bf 5898},  201--211 (2005).

\end{thebibliography}
\bibliographystyle{spiebib}

\end{document}